# Cross Domain Few-Shot Class-Incremental Audio Classification Via Adversarial Contrastive Learning

*Yongjie Si, Yanxiong Li*[*]*, Sen Huang, Beibei Liu*

School of Electronic and Information Engineering, South China University of Technology, Guangzhou, China
eeyongjiesi@mail.scut.edu.cn, eeyxli@scut.edu.cn, huangsen@scut.edu.cn, eebbliu@scut.edu.cn

## Abstract

Current Few-shot Class-incremental Audio Classification (FCAC) methods assume that samples of base and incremental classes are in the same domain (following the same distribution). However, there is generally a domain shift between the above two types of samples. In this paper, we explore the problem of Cross Domain FCAC where samples of base and incremental classes have domain shift. We propose a strategy of adversarial contrastive training which enables the model to effectively classify samples of different classes from unseen domains. The model consists of an encoder and a classifier. The encoder is trained in base session but frozen in incremental sessions, whereas the classifier is trained in all sessions. Experiments are done on six pairs of cross-domain datasets. Results show that our method exceeds state-of-the-art methods in average accuracy. The code is at https://github.com/YongjieSi/ACL.

## 1. Introduction

Audio classification (AC) is a task to identify different audio classes. It has widespread applications, such as wildlife protection [1], road surveillance [2], classification of acoustic scene [3], video analysis [4] and speaker analysis [5]. Many AC methods [6]-[9] require abundant samples of predefined classes for model training. The trained model generally lacks the ability to recognize new classes. In real applications, the number of classes is not fixed but dynamically changes and the number of training samples per class is generally limited. To address challenges such as the variable number of classes and scarcity of training samples per class, Few-shot Class-incremental Audio Classification (FCAC) methods are proposed [10]-[16]. They can continually identify new classes using a few training samples per class and meanwhile memorize old ones. A common practice in the FCAC methods is to decompose the model into an encoder and a classifier. Existing FCAC methods have different technical characteristics, such as Dynamic Few-Shot Learning (DFSL) [10], prototype-based classifier [11]-[13], stochastic classifier [14], linear classifier [15], [16], and ridge regression classifier [17].

The above efforts have contributed to the development of AC, but they still leave something to be improved. For example, these methods assume that the samples of different classes in both base and incremental sessions are from the same domain. However, there is a domain shift between the samples from base and incremental sessions in many applications. In this paper, we tackle a new and practical problem in which domain shift and class increment coexist in different session. This problem is called Cross Domain FCAC (CD-FCAC). It is an extension of the FCAC problem which does not consider the domain shift of samples from different sessions. To our best knowledge, the work in this paper is the first effort to solve the CD-FCAC problem. The motivation for doing this work is due to the existence of CD-FCAC problems in many practical applications. For example, in smart home scenarios, the intelligent agents need to incrementally recognize different audio classes (such as voice commands, environmental sounds, and musical sounds) from different domains (e.g., different datasets).

In addition to the catastrophic forgetting of old classes in incremental learning, there are two key challenges in the CD-FCAC. First, there is domain shift between samples of both base and incremental classes. Second, there is label difference between base and incremental classes. To obtain high accuracy, the model needs to have strong ability to overcome the impacts of class increment and domain shift. To address the above two challenges, we propose a strategy of adversarial contrastive training which adopts both adversarial training and contrastive loss. To alleviate domain shift, diverse adversarial samples are generated to simulate pseudo-target domains. They and the source-domain samples are used together to train the model for extracting domain-invariant embeddings. To enable the model to generalize well to unseen classes, we use the loss of supervised contrastive learning for training the model in the base session. This loss aims to aggregate the embeddings of samples of the same class while keeping the embeddings of samples of different classes away from each other. Hence, the embeddings are expected to be intra-class compact and inter-class separable, which allows the classifier to establish robust decision boundaries based on a few samples of new classes. To prevent the model from forgetting the knowledge of old classes, the encoder is frozen and mean vectors of old classes' embeddings are saved in each incremental session, while the classifier is updated for learning new classes in each session.

Experiments are done on six pairs of cross-domain datasets from three public datasets of the LS-100, NSynth-100, and FSC-89 [17]. Results show that our method exceeds previous methods in average accuracy. The contributions of the work are summarized as follows.

1) We explore a new problem of CD-FCAC where the domain shift and class increment coexist.

2) We propose an adversarial contrastive training strategy which is a fusion of adversarial training and contrastive learning.

3) We propose a CD-FCAC method which exceeds state-of-the-art methods in average accuracy.

## 2. Method

### 2.1. Problem Definition

The CD-FCAC consists of $M$ sessions, including one base session (session 0) and ($M$-1) incremental sessions (sessions 1 to ($M$-1)). we consider a user-friendly setting where the samples in base session are from a single source domain, while the samples in incremental sessions are from another domain. The base classes (in the base session) and incremental classes (in the incremental sessions) are called source-domain classes and target-domain classes, respectively. In addition, audio classes in the prior and current sessions are called old and new classes, respectively. The training and testing datasets of different sessions are $\{\boldsymbol{D}_0^{tr}, \boldsymbol{D}_1^{tr}, ..., \boldsymbol{D}_m^{tr}, ..., \boldsymbol{D}_{M-1}^{tr}\}$ and $\{\boldsymbol{D}_0^{te}, \boldsymbol{D}_1^{te}, ..., \boldsymbol{D}_m^{te}, ..., \boldsymbol{D}_{M-1}^{te}\}$, respectively. $\boldsymbol{D}_m^{tr}$ and $\boldsymbol{D}_m^{te}$ have the same label set $\boldsymbol{L}_m$. The datasets in various sessions don't have the same classes, namely $\forall m, h$ and $m \neq h$, $\boldsymbol{L}_m \cap \boldsymbol{L}_h = \phi$. In the $m$th session, only $\boldsymbol{D}_m^{tr}$ can be used to train the model, and the trained model is tested on the testing datasets of current and all previous sessions, namely $\boldsymbol{D}_0^{te} \cup \boldsymbol{D}_1^{te} ... \cup \boldsymbol{D}_m^{te}$. The dataset $\boldsymbol{D}_0^{tr}$ has many classes and each class has sufficient training samples, while the dataset $\boldsymbol{D}_m^{tr}$ ($1 \leq m \leq (M\text{-}1)$) only has $N$ classes and each class only has $K$ training samples.

* Corresponding author: Yanxiong Li (eeyxli@scut.edu.cn).

## 2.2. Method Framework

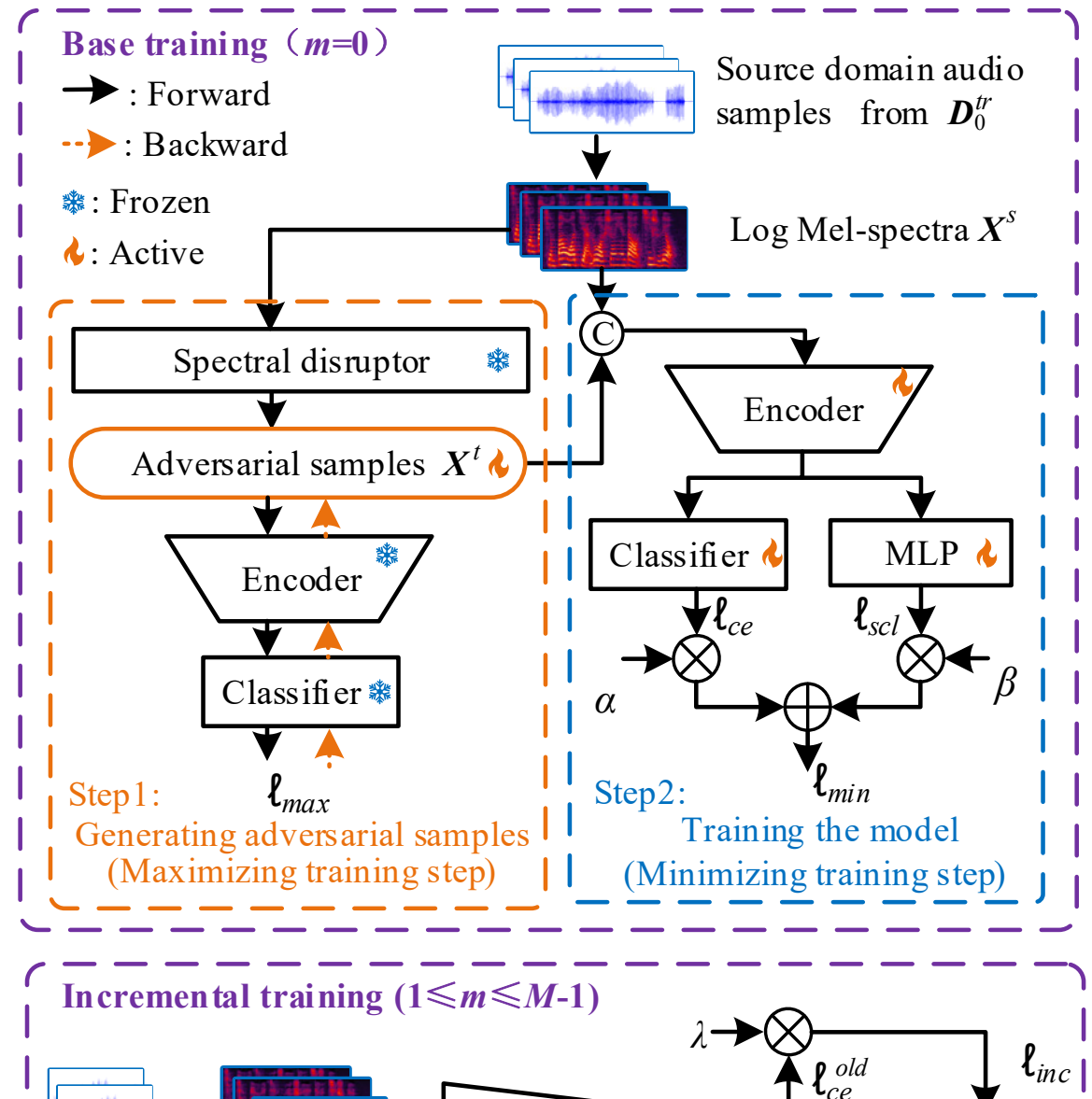

Figure 1: *Framework of our method. In base training, we learn domain-invariant embeddings via adversarial contrastive training using a spectral disruptor. In incremental training, we freeze the encoder and update the classifier with embeddings of new classes samples and mean vectors of old classes' embeddings.* $\alpha, \beta, \lambda$: *three adjustable coefficients;* $\ell_{ce}^{new}$: *cross-entropy loss* $\ell_{ce}$ *on embeddings of new classes;* $\ell_{ce}^{old}$: $\ell_{ce}$ *on mean vectors of old classes' embeddings;* MLP: *multi-layer perceptron;* $\boldsymbol{\mu}_{y_i}$: *mean vectors of embeddings for class* $y_i$; $\boldsymbol{L}_m$: *label set in session m.*

Figure 1 shows our method's framework which includes base ($m$=0) and incremental ($1 \leq m \leq (M\text{-}1)$) training sessions. The model consists of an encoder and a classifier. The log Mel-spectrogram of audio samples are used as the input of the model. To increase the diversity of adversarial samples, we design a spectral disruptor used in the adversarial contrastive training to perturb the time-frequency elements of the log Mel-spectrogram. The spectral disruptor includes two random convolutional layers with the same parameters [31], which is initialized with normal distribution and is frozen in the generation of adversarial samples. The kernel size is selected randomly from a candidate pool. The stride and padding are set to preserve the shape of log Mel-spectrogram. In the base session, the model is trained by an adversarial contrastive training strategy for extracting domain-invariant embeddings. In each incremental session, the encoder is frozen to preserve the knowledge of base classes, while the classifier is updated to enable the model to recognize new classes. In inference, cosine distance between embeddings and classifier weights is used to identify each test sample.

In the base training session, each training episode includes two steps, i.e., generating adversarial samples and training the model. In the first step, the model is frozen and the adversarial samples are generated based on the single source domain samples by maximizing the loss $\ell_{max}$ (defined by Eq. 7). In the second step, the model is trained by minimizing the loss $\ell_{min}$ (defined by Eq. 8) using both the generated adversarial samples and source domain samples. To mitigate the model's forgetting of old classes, the mean vector of embeddings for class $y_i$ ($y_i \in \boldsymbol{L}_0 \cup \cdots \boldsymbol{L}_m$) is saved and computed by

$$\boldsymbol{\mu}_{y_i} = \frac{1}{K}\sum_{k=1}^{K} f_E(\boldsymbol{x}_k) , \qquad (1)$$

where $K$ denotes the number of support samples per class; $f_E(\boldsymbol{x}_k)$ denotes the embedding of sample $\boldsymbol{x}_k$.

In the $m$th ($1 \leq m$) incremental training session, the classifier is updated using mean vectors of old classes' embeddings and the embeddings extracted from the samples in $\boldsymbol{D}_m^{tr}$ by minimizing the incremental loss $\ell_{inc}$ which is defined by

$$\ell_{inc} = \ell_{ce}^{new} + \lambda \ell_{ce}^{old}, \qquad (2)$$

where $\lambda$, $\ell_{ce}^{new}$ and $\ell_{ce}^{old}$ denote an adjustable coefficient, the cross-entropy loss $\ell_{ce}$ on the training data of the $m$th session and the $\ell_{ce}$ on the mean vectors of old classes' embeddings, respectively. $\ell_{ce}$ is defined by

$$\ell_{ce}((\boldsymbol{x}_i, y_i);\ \theta) = -\,log \frac{\exp\left(\eta \cdot \cos(f_E(\boldsymbol{x}_i), \boldsymbol{c}_{y_i})\right)}{\sum_{y_j=1}^{N} \exp\left(\eta \cdot \cos(f_E(\boldsymbol{x}_i), \boldsymbol{c}_{y_j})\right)}, \qquad (3)$$

where $\theta = \{E, C\}$ represents the model, consisting of an encoder $E$ and a classifier $C$. $\boldsymbol{c}_{y_i}$ is the weight vector corresponding to the $j$-th class. $\eta$, $\exp(\cdot)$ and $\cos(\cdot,\cdot)$ are a scale factor, exponential and cosine functions, respectively.

## 2.3. Adversarial Contrastive Training

An adversarial contrastive training strategy is proposed to make the learned embeddings domain-invariant, intra-class compact, and inter-class separable. Building upon the framework of adversarial training [19]-[22], the problem of single source-domain generalization can be formulated by

$$\min_{\theta} \sup_{D_t} \{\mathbb{E}[\ell(D_s;\ \theta) : d(D_t, D_s) \leq \rho]\}, \qquad (4)$$

where $D_s$ and $D_t$ are the distributions of source and target domains, respectively. $\rho$ is the maximum distance constraint between $D_s$ and $D_t$. $d(D_t, D_s)$ denotes the distance between $D_s$ and $D_t$, and is defined by

$$d(D_t, D_s) = \|f_E(\boldsymbol{x}^s) - f_E(\boldsymbol{x}^t)\|_2^2 + \infty \cdot 1\{y^t \neq y^s\}, \qquad (5)$$

where $\boldsymbol{x}^s$ (or $\boldsymbol{x}^t$) and $y^s$ (or $y^t$) are the samples and corresponding labels from source (or target) domains, respectively. Eq. 5 defines the distance metric as Euclidean for samples with the same label and infinity for those with different labels. This ensures semantic consistency between source domain samples and the generated adversarial samples. The solution to Eq. 4 guarantees good performance of the model $\theta$ for all target distributions $D_t$ satisfying $d(D_t, D_s) \leq \rho$ [20]. Specifically, the $\sup\{\cdot\}$ problem serves to identify the worst-case target distribution $D_t$ by maximizing the expectation of loss $\ell$, namely $\mathbb{E}[\ell]$. The $\min\{\cdot\}$ problem optimizes the model's parameters $\theta$ by minimizing $\mathbb{E}[\ell]$. After iterations of optimization, the model learns embeddings that are robust to all distributions bounded by $\rho$. The Lagrangian relaxation derived from Eq. 4 with a penalty $\gamma \geq 0$ is used here, namely

$$\min_{\theta} \sup_{D_t} \{\mathbb{E}[\ell(D_t;\ \theta)] - \gamma d(D_t, D_s)\}. \quad (6)$$

Following [19], the problem in Eq. 6 can be solved through alternating maximizing and minimizing training steps.

In the maximizing training step, the model is frozen and the adversarial samples are generated by maximizing $\ell_{max}$

$$\ell_{max} = \ell_{ce}((\boldsymbol{x}^t, y^s);\ \theta) - \gamma d\big((\boldsymbol{x}^t,\ y^s), (\boldsymbol{x}^s,\ y^s)\big). \quad (7)$$

In the minimizing training step, the adversarial samples generated in the maximizing step and the source domain samples are used to train the model by minimizing loss $\ell_{min}$. To mitigating few-shot overfitting, we integrate the supervised contrastive loss [18] into $\ell_{min}$. This reinforces intra-class compactness and inter-class separability, pre-structuring a discriminative embedding space that facilitates the establishment of robust decision boundaries for new classes with a few samples. Accordingly, $\ell_{min}$ is defined by

$$\ell_{min} = \alpha\ell_{ce} + \beta\ell_{scl}, \quad (8)$$

where $\alpha$ and $\beta$ are two adjustable coefficients; $\ell_{scl}$ is defined by

$$\ell_{scl} = \sum_{b \epsilon B} \frac{-1}{|I(b)|} \sum_{i \epsilon I(b)} log \frac{exp(\tau f_E'(x_b) \cdot f_E'(x_i))}{\sum_{j=A(b)} exp(\tau f_E'(x_b) \cdot f_E'(x_j))}, \quad (9)$$

where $B$, $\tau$, $\boldsymbol{x}_b$, and $\boldsymbol{x}_i$ denote the index set of samples in a training episode, a scale factor, anchor sample, and positive sample of the same class as $\boldsymbol{x}_b$, respectively. $f_E'(\cdot)$ is the embedding of $f_E(\cdot)$ output by the MLP. For any index $b$, $I(b) = \{i \epsilon A(b) | y_i = y_b\}$ stands for the set of indices that share the same label as $\boldsymbol{x}_b$. $A(b) = B \backslash \{b\}$ denotes the index set except $b$. The process of the adversarial contrastive training strategy is summarized as Algorithm 1, as shown in Table 1.

Table 1: *Algorithm* 1 *Adversarial contrastive training.*

**Input**: training dataset $\boldsymbol{D}_0^{tr}$ from source domain;
**Require**: learning rate $\zeta_{max}$ and $\zeta_{min}$; epoch number *R*
**Output**: the trained model $\theta$ in base session
1. **While** minimizing training **do**
2. Select audio samples from $\boldsymbol{D}_0^{tr}$ and extract its log-Mel spectrogram $(\boldsymbol{X}^s,\ \boldsymbol{Y}^s)$
3. **While** maximizing training **do**
4. $\boldsymbol{X}^t$ ←Spectral disturbance on $\boldsymbol{X}^s$
5. **For** *R* epochs **do**
6. $\boldsymbol{X}^t \leftarrow \boldsymbol{X}^t + \zeta_{max} \nabla_{\boldsymbol{X}^t}(\ell_{ce}((\boldsymbol{X}^t, \boldsymbol{Y}^s);\ \theta) - \gamma d((\boldsymbol{X}^t,\ \boldsymbol{Y}^s), (\boldsymbol{X}^s,\ \boldsymbol{Y}^s)))$
7. **End For**
8. **End While**
9. $\boldsymbol{X} = [\boldsymbol{X}^s, \boldsymbol{X}^t],\ \boldsymbol{Y} = [\boldsymbol{Y}^s, \boldsymbol{Y}^s]$
10. $\theta \leftarrow \theta - \zeta_{min} \nabla_\theta \ell_{min}(\boldsymbol{X}, \boldsymbol{Y}; \theta)$
11. **End While**

# 3. Experiments

## 3.1. Experimental Datasets

Table 2: *Detailed information of the* LS-*100/NSynth -100/FSC-89.*

| Param | $\boldsymbol{D}_0$ | | $\boldsymbol{D}_m$ $(1 \leq m \leq (M\text{-}1))$ | |
|---|---|---|---|---|
| | $\boldsymbol{D}_0^{tr}$ | $\boldsymbol{D}_0^{te}$ | $\boldsymbol{D}_m^{tr}$ | $\boldsymbol{D}_m^{te}$ |
| #C | 60/55/59 | 60/55/59 | 40/45/30 | 40/45/30 |
| L(h) | 16.66/12.23 /13.11 | 3.33/1.52 /3.28 | 11.11/5.00 /4.17 | 2.22/5.00 /1.67 |
| #S/C | 500/200/800 | 100/100/200 | 500/100/500 | 100/100/200 |

#C: number of classes, L(h): Length(hours) #S/C: number of samples per class.

Experimental data are from three public datasets, including the LS-100, NSynth-100 and FSC-89 [17]. They are widely used in prior work, and released at the three websites[1]. Each dataset is split into *M* parts, namely $\boldsymbol{D}_m$ $(0 \leq m \leq (M\text{-}1))$, without overlaps of classes. $\boldsymbol{D}_m$ consists of training dataset $\boldsymbol{D}_m^{tr}$ and testing dataset $\boldsymbol{D}_m^{te}$. Table 2 lists the details of the three datasets, in which each digit X, Y and Z in the X/Y/Z represent the contents for the LS-100, NSynth-100 and FSC-89, respectively. For example, 60/55/59 denotes that the number of classes for the LS-100, NSynth-100 and FSC-89 is 60, 55 and 59, respectively.

These three datasets are adopted to construct six pairs of cross-domain datasets, namely FS→NS, FS→LS, NS→FS, NS→LS, LS→FS and LS→NS，where LS, NS and FS denote LS_100, NS-100 and FSC_89, respectively. For example, in the FS→NS, the item on the left side of the arrow (i.e., FS) denotes the source domain data (namely $\boldsymbol{D}_0$ of FSC-89), and the item on the right side of the arrow (i.e., NS) denotes the target domain data (namely $\boldsymbol{D}_m$ $(1 \leq m \leq (M\text{-}1))$ of NSynth-100).

The usage of the above cross-domain datasets is based on two considerations. First, we need to address the problems of both domain shift (cross-domain) and class increment (cross-class) in this paper. The FSC-89, NSynth-100 and LS-100 have different audio classes and are recorded in different environments using various recording devices. Hence, there are both domain shift and class increment between the above three datasets. Second, the construction of the above cross-domain datasets meets the requirements of practical applications. For example, in smart home application scenarios, the humanoid robots deployed with audio classification models are required to incrementally recognize voice commands (speech dataset), environmental sound (sound-event dataset), and musical sounds (musical-instrument dataset) from different domains.

## 3.2. Experimental Setup

Average Accuracy (AA) is adopted to measure the overall performance of various methods in all sessions and defined by

$$AA = \frac{1}{M} \sum_{m=0}^{M-1} A_m \ , \quad (10)$$

where $A_m$ denotes the accuracy in session *m*.

Following the practice of [10]-[15], we use ResNet-18 [30] as the encoder. In the base session, the model is trained for 200 epochs. Each epoch has 50 episodic training steps. In each episodic training step, *NK* support samples and *NQ* query samples are randomly chosen from $\boldsymbol{D}_m^{tr}$. In each incremental session, the model is trained for 100 epochs. All testing datasets of seen classes are used to evaluate the model. Dimensions of log Mel- spectrogram and embeddings are set to 128 and 512, respectively. Coefficients $\alpha$, $\beta$, $\gamma$, $\lambda$, and learning rate are set to 1, 0.2, 1, 0.6, and 0.1, respectively.

## 3.3. Ablation Experiments

Table 3: *Results obtained* by *different methods on* NS→LS.

| Case | SC | AT | Accuracy scores in various sessions (%) | | | | | | | | | AA (%) |
|---|---|---|---|---|---|---|---|---|---|---|---|---|
| | | | 0 | 1 | 2 | 3 | 4 | 5 | 6 | 7 | 8 | |
| ① | × | × | 99.96 | 88.97 | 83.45 | 78.21 | 73.27 | 68.99 | 65.43 | 62.01 | 58.84 | 75.46 |
| ② | √ | × | **100.00** | 91.75 | 85.15 | 79.28 | 74.32 | 69.97 | 66.10 | 62.58 | 59.36 | 76.50 |
| ③ | × | √ | 99.94 | 92.40 | 86.73 | 80.98 | 75.74 | 71.28 | 67.99 | 64.39 | 61.13 | 77.84 |
| ④ | √ | √ | 99.93 | **94.13** | **88.38** | **82.52** | **77.15** | **72.85** | **69.17** | **65.43** | **62.22** | **79.09** |

SC: Supervised Contrastive loss; AT: Adversarial Training.

This section discusses the contributions of Supervised Contrastive loss (SC) and Adversarial Training (AT) to the performance of our method. Without the loss of generality, the NS→LS is used in ablation and extended experiments. Table 3 lists the results of our method with various combinations of SC and AT on the NS→LS. In case ①, the model is trained by minimizing $\ell_{ce}$ without SC and AT in the base session. In cases ② and ③, the model is trained by minimizing $\ell_{ce}$ using SC and AT only, respectively. Based on the results in all four cases, it can be obtained that the SC and AT contribute to the performance gain. When both of them are used (case ④), our method obtains the highest AA scores of 79.09%.

### 3.4. Comparison of Different Methods

Our method is compared to nine baseline methods, including the DFSL [10], CEC [27], PAN [11], AMFO [15], AMFO+TSA, AMFO+LDP, AMFO+AFA, AMFO+AMTF and PCR [28]. Among them, the DFSL, CEC, PAN and AMFO are the FCAC methods without considering cross domain. The AMFO+TSA, AMFO+LDP, AMFO+AFA and AMFO+AMTF are the combination of both few-shot cross-domain methods (namely TSA [25], LDP [26], AFA [21] and AMTF [22]) and the FCAC method (namely AMFO). PCR is a method of cross-domain few-shot class-incremental point cloud recognition, where the problem setting is the same to ours.

The DFSL method designs an attentive weight generator for producing classifier weight vectors to adapt to new classes. The CEC method designs a graph model to propagate contextual information between classifiers. The PAN method designs self-attention modified prototypes to expand classifier. The AMFO method adopts discriminative and generalizable classifier and embeddings. The TSA method uses lightweight parametric adapters, allowing task-specific adaptation. The LDP method uses local and global distillation prototype network to improve the generalizability of embeddings. The AFA method adopts the adversarial feature augmentation to bridge the domain gap in few-shot learning. The AMTF method adopts a meta-training strategy to improve the generalization ability of existing methods in few-shot learning. The PCR method distinguishes between base classes and new classes from an out-of-distribution detection perspective.

Following the practice of [10]-[16], the value of ($N$, $K$) is typically set to (5, 5). The AA scores of both source and target domain classes obtained by different methods on the six pairs of cross-domain datasets are listed in Table 4. Our method achieves AA scores of 46.89%, 41.67% 85.17%, 79.09%, 80.05% and 79.78% on the FS→NS, FS→LS, NS→FS, NS→LS, LS→FS and LS→NS, respectively. The above AA scores are higher than the counterparts obtained by other methods. The advantage of our method in AA is mainly attributed to the adversarial contrastive training strategy. The strategy enables the model to extract domain-invariant embeddings with intra-class compactness and inter-class separability. Thus, our model has strong generalization ability for unseen domains and classes.

Table 4: *Average* accuracy *obtained by various methods (in %).*

| Methods | FS→NS | FS→LS | NS→FS | NS→LS | LS→FS | LS→NS |
|---|---|---|---|---|---|---|
| DFSL | 37.21 | 37.54 | 82.58 | 76.62 | 77.83 | 75.30 |
| CEC | 40.75 | 39.29 | 83.75 | 77.31 | 78.57 | 76.86 |
| PAN | 40.88 | 39.35 | 83.83 | 77.50 | 78.71 | 77.34 |
| AMFO | 45.31 | 37.05 | 83.86 | 76.93 | 78.10 | 71.92 |
| AMFO+TSA | 44.13 | 36.24 | 84.19 | 77.24 | 78.21 | 78.20 |
| AMFO+LDP | 43.30 | 33.61 | 84.70 | 77.40 | 76.35 | 72.60 |
| AMFO+AFA | 44.75 | 41.52 | 84.46 | 78.50 | 79.47 | 78.68 |
| AMFO+AMTF | 44.29 | 40.76 | 83.93 | 77.95 | 79.11 | 78.21 |
| PCR | 38.70 | 36.98 | 75.61 | 78.22 | 78.79 | 78.80 |
| Ours | 46.89 | 41.67 | 85.17 | 79.09 | 80.05 | 79.78 |

In addition, we perform two extended experiments. In the first extended experiment, we discuss the impacts of $N$ and $K$ values on the AA scores obtained by our method. Figure 2 depicts the results obtained by our method on the NS→LS when $N$-way $K$-shot is set to different values. When $N$ is fixed, the AA score steadily increases as $K$ grows. The reason for the above result is that the larger the $K$ value, the more training samples for each class, and the better the model performance. When $K$ is fixed, the AA scores reach the maximum when $N$ equals 5. If $N$ decreases below 5, the number of sessions increases and then the model easily forgets old classes. Therefore, the AA scores will decrease. If $N$ increases over 5, the number of classes per session increases. Hence, heavier confusions of different classes will occur in each session and thus the AA scores will decline accordingly.

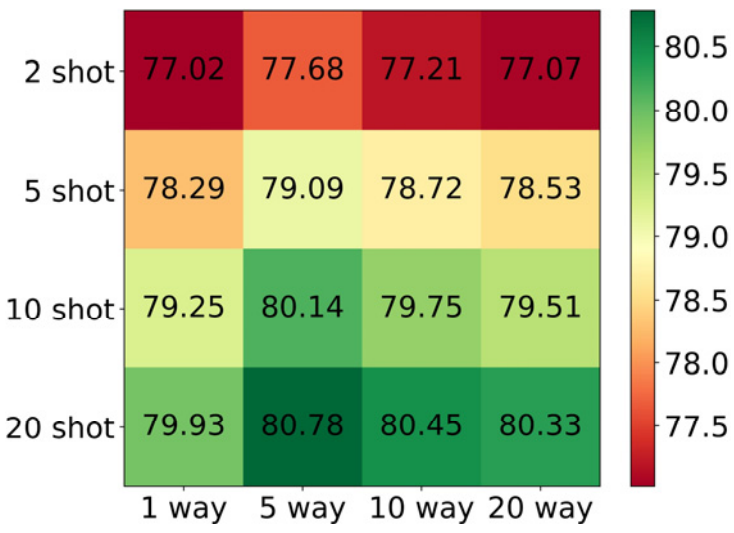

Figure 2: *AA scores (in %) obtained by our method on the NS→LS using various values of N-way K-shot.*

In the second extended experiment, we use t-SNE [29] to show the effectiveness of the adversarial contrastive training strategy. Figure 3 visualizes the testing embeddings without using or using the proposed strategy on the NS→LS, where five source and target classes are randomly selected from the NSynth-100 and LS-100, respectively. As shown in Fig. 3 (a), without using the adversarial contrastive training strategy, the embeddings of most target domain classes overlap each other, and they also overlap with the embeddings of source-domain classes. The reason is that the encoder trained solely on source domain data fails to extract discriminative embeddings for the target domain. Fig. 3 (b) shows that our method using the proposed strategy makes the testing embeddings of various classes distribute in different places with almost no overlap.

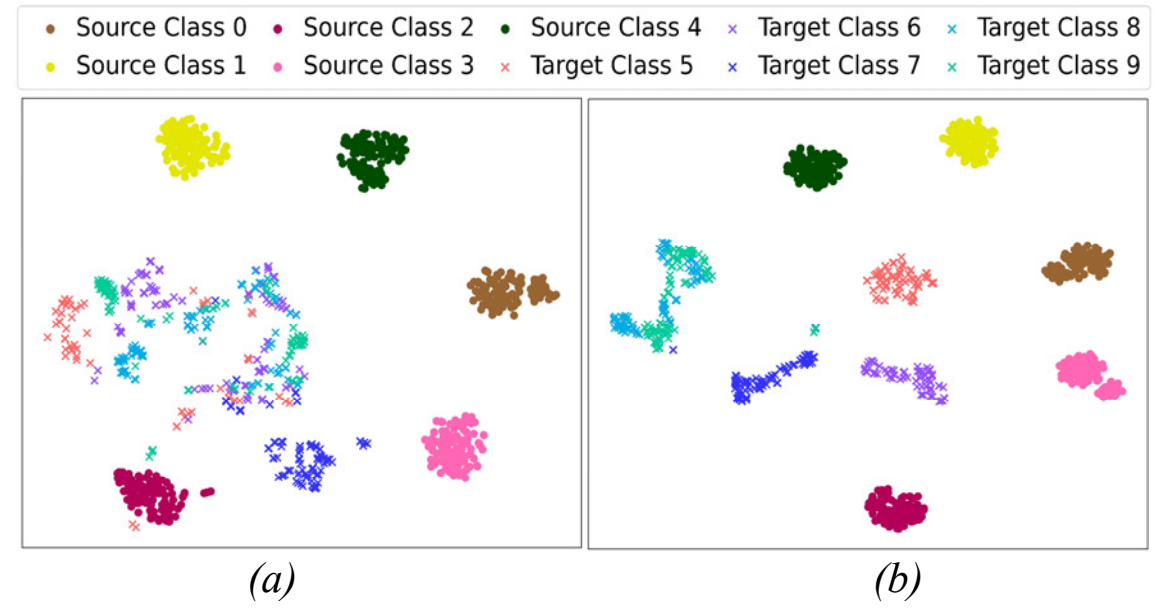

Figure 3: *Distributions of testing embeddings of five source and target classes on the NS→LS (a) without using and (b) using the proposed adversarial contrastive training strategy.*

## 4. Conclusions

We solved the CD-FCAC problem via the adversarial training and supervised contrastive loss. Based on the description of our method and results, we can draw two conclusions. First, our method exceeds state-of-the-art methods in average accuracy. Second, the proposed adversarial contrastive training strategy benefits to improve the model's performance. In next work, we will enhance the accuracy by designing more effective encoder and classifier.

[1] https://www.modelscope.cn/datasets/pp199124903/LS-100/summary
https://www.modelscope.cn/datasets/pp199124903/FSC-89/summary
https://www.modelscope.cn/datasets/pp199124903/NSynth-100/summary

## 5. Acknowledgements

This work was supported by the national natural science foundation of China (62371195, 62111530145, 61771200), the exchange project of the 10th Meeting of the China-Croatia Science and Technology Cooperation Committee (No. 10-34).

## 6. Generative AI Use Disclosure

No generative AI tools (e.g., LLM/ChatGPT) were used to produce any part of this work, except that a limited amount of grammar/spelling checking may have been performed on the final text using standard tools. The authors take full responsibility for the content.

## 7. References

[1] A. Terni, N. Ortolani, I. Nolasco, E. Bonitos, and S. Cecchi, "Comparison of feature extraction methods for sound-based classification of honey bee activity," *IEEE/ACM TASLP*, vol. 30, pp. 112-122, 2022.

[2] Y. Li, X. Li, Y. Zhang, M. Liu, and W. Wang, "Anomalous sound detection using deep audio representation and a BLSTM network for audio surveillance of roads," *IEEE Access*, vol. 6, pp. 58043-58055, 2018.

[3] Y. Li, J. Tan, G. Chen, J. Li, Y. Si, and Q. He, "Low-complexity acoustic scene classification using parallel attention-convolution network," in *Proc. of Interspeech*, 2024, pp. 1-5.

[4] W. Pang, et al., "Detecting video anomalies by jointly utilizing appearance and skeleton information," *Expert Syst. Appl.*, vol. 246, pp. 1-12, 2024.

[5] Y. Li, W. Wang, M. Liu, Z. Jiang, and Q. He, "Speaker clustering by co-optimizing deep representation learning and cluster estimation," *IEEE TMM*, vol. 23, pp. 3377-3387, 2021.

[6] Y.-X. Lin, J.-C. Lin, W.-L. Wei and J.-C. Wang, "Learnable counterfactual attention for music classification," *IEEE TASLP*, vol. 33, pp. 570-585, 2025.

[7] M. Agarwal, K. S. Gill, S. Chattopadhyay and M. Singh, "Classification of urban sound using sequential convolutional neural network model and its visualisation," in *Proc. of ICITEICSI*, 2024, pp. 1-5.

[8] K. M. Lim, C. P. Lee, Z. Y. Lee, J. Y. Lim and J. N. Mogan, "Acoustic event classification with enhanced efficientNet," in *Proc. of ICSPC*, 2024, pp. 13-17.

[9] S. Barahona, D. de Benito-Gorrón, D. T. Toledano and D. Ramos, "Enhancing conformer-based sound event detection using frequency dynamic convolutions and beats audio embeddings," *IEEE/ACM TASLP*, vol. 32, pp. 3896-3907, 2024.

[10] Y. Wang, N. J. Bryan, M. Cartwright, J. P. Bello, and J. Salomon, "Few shot continual learning for audio classification," in *Proc. of IEEE ICASSP*, 2021, pp. 321-325.

[11] Y. Li, W. Cao, W. Xian, J. Li, and E. Benito's, "Few-shot class-incremental audio classification using dynamically expanded classifier with self-attention modified prototypes," *IEEE TMM*, vol.26, pp.1346-1360, 2024.

[12] W. Xie, et al., "Few-shot class-incremental audio classification via discriminative prototype learning," *Expert Syst. Appl.*, vol. 225, 2023, Art. no. 120044.

[13] W. Xie, et al., "Few-shot class-incremental audio classification using adaptively-refined prototypes," in *Proc. of Interspeech*, 2023, pp. 301-305.

[14] Y. Li, W. Cao, J. Li, W. Xie, and Q. He, "Few-shot class-incremental audio classification using stochastic classifier," in *Proc. of Interspeech*, 2023, pp. 4174-4178.

[15] Y. Li, J. Li, Y. Si, J. Tan, and Q. He, "Few-shot class-incremental audio classification with adaptive mitigation of forgetting and overfitting," *IEEE/ACM TASLP*, vol. 32, pp. 2297-2311, 2024.

[16] Y. Si, Y. Li, J. Tan, G. Chen, Q. Li, and M. Russo, "Fully few-shot class-incremental audio classification with adaptive improvement of stability and plasticity," *IEEE TASLP*, vol. 33, pp. 418-433, 2025.

[17] Y. Si, Y. Li, J. Tan, Q. He, and I. Kwak, "Fully few-shot class-incremental audio classification using multi-level embedding extractor and ridge regression classifier," in *Proc. of Interspeech*, 2025, pp. 1318-1322.

[18] P. Khosla, P. Teterwak, C. Wang, A. Sarna, Y. Tian, P. Isola, A. Maschinot, C. Liu, and D. Krishnan, "Supervised contrastive learning," in *Proc. of NIPS*, 2020. pp. 18661-18673.

[19] A. Sinha, H. Namkoong, and J. Duchi, "Certifying some distributional robustness with principled adversarial training," in *Proc. of International Conference on Learning Representations*, 2018, pp. 1-34.

[20] R. Volpi, H. Namkoong, O. Sener, and J.C. Duchi, V. Murino, and S. Savarese, "Generalizing to unseen domains via adversarial data augmentation," in *Proc. of Advances in Neural Information Processing Systems*, 2018, pp. 1-11.

[21] Y. Hu, A.J. Ma, "Adversarial feature augmentation for cross-domain few-shot classification," in *Proc. of ECCV*, 2022, pp. 20-37.

[22] P. Tian, and S. Xie, "An adversarial meta-training framework for cross-domain few-shot learning," *IEEE Transactions on Multimedia*, vol. 25, pp. 6881-6891, 2023.

[23] K. Lee, K. Lee, J. Shin, and H. Lee, "Network randomization: A simple technique for generalization in deep reinforcement learning," in *Proc. of ICLR*, 2020, pp. 1-22.

[24] X. Glorot, and Y. Bengio, "Understanding the difficulty of training deep feedforward neural networks," in *Proc. of the thirteenth international conference on artificial intelligence and statistics*, 2010, pp. 249-256.

[25] W. Li, X. Liu and H. Bilen, "Cross-domain Few-shot Learning with Task-specific Adapters," in *Proc. of CVPR*, 2022, pp. 7151-7160,

[26] F. Zhou, P. Wang, L. Zhang, W. Wei and Y. Zhang, "Revisiting Prototypical Network for Cross Domain Few-Shot Learning," in *Proc. Of* CVPR, 2023, pp. 20061-20070.

[27] C. Zhang, N. Song, G. Lin, Y. Zheng, P. Pan, and Y. Xu, "Few-shot incremental learning with continually evolved classifiers," in *Proc. of CVPR*, 2021, pp. 12450 -1245.

[28] Y. Tan, and X. Xiang, "Cross-domain few-shot incremental learning for point-cloud recognition," in *Proc. of IEEE/CVF Winter Conference on Applications of Computer Vision*, 2024, pp. 2296-2305.

[29] L. V. D. Maaten, and G. Hinton, "Visualizing data using T-SNE," *J. Mach. Learn. Res.*, vol. 9, no. 96, pp. 2579-2605, 2008.

[30] K. He, X. Zhang, S. Ren and J. Sun, "Deep Residual Learning for Image Recognition," in *Proc. of IEEE CVPR*, 2016, 1022 pp. 770-778.

[31] Kimin Lee, Kibok Lee, Jinwoo Shin, and Honglak Lee. Network randomization: A simple technique for generalization in deep reinforcement learning. in *Proc. of IEEE ICLR*, 2020.